\documentclass[preprint]{revtex4-1}
\usepackage{graphicx}
\usepackage{xcolor}
\usepackage{hyperref}

\usepackage{amsmath}
\usepackage{SIunits}
\newcommand{\wstar}{-0.023}
\newcommand{\Qstar}{0.25}

\begin{document}
\title{Roles of icosahedral and crystal-like order \\ in hard spheres glass transition}

\author{Mathieu Leocmach} 

\author{Hajime Tanaka}
\email{tanaka@iis.u-tokyo.ac.jp}
\affiliation{ {Institute of Industrial Science, University of Tokyo, 4-6-1 Komaba, Meguro-ku, Tokyo 153-8505, Japan} }

\date{Received September 27, 2011}

\begin{abstract}
\textbf{
A link between structural ordering and slow dynamics has recently attracted much attention from the context of the origin of glassy slow dynamics. Candidates for such structural order are icosahedral, exotic amorphous, and crystal-like. Each type of order is linked to a different scenario of glass transition. Here we experimentally access local structural order in polydisperse hard spheres by particle-level confocal microscopy. We identify the key structures as icosahedral and \textmd{\textsc{fcc}}-like order, both statistically associated with slow particles. However, when approaching the glass transition, the icosahedral order does not grow in size whereas crystal-like order grows. It is the latter that governs the dynamics and is linked to dynamic heterogeneity. This questions the direct role of the local icosahedral ordering in glassy slow dynamics and suggests that the growing lengthscale of structural order is essential for the slowing down of dynamics and the nonlocal cooperativity in particle motion. 
}
\end{abstract}
\maketitle

\section*{}

Upon cooling, a liquid usually undergoes a first order transition to an ordered ground state (crystal or quasi-crystal). However, it is generally possible to avoid the transition and to form a metastable supercooled liquid. Upon further cooling the dynamics slows down dramatically by many orders of magnitude, leading to the impossibility to equilibrate the system in the experimental time scale below a temperature $T_g$ named the glass transition temperature. The nature of the glass transition (thermodynamic or kinetic, structural or purely dynamical) is still a matter of debate after decades of intensive research (see~\citep{cavagna2009supercooled,BerthierR} for comprehensive reviews).

A part of the answer is believed to come from the dynamics in itself: a supercooled liquid is dynamically heterogeneous (see~\citep{BerthierR} for a review) and the characteristic size of the dynamical heterogeneity is growing when approaching the glass transition~\citep{yamamoto1998,Donati1999a} (although non-monotonic behaviour has been observed recently~\cite{Kob2011}). The dynamical arrest may then be the analogue of the slowing down observed near a critical point when the characteristic size of the fluctuations diverges, although slowing down at a particle level is absent in ordinary critical phenomena. Here we note that the lengthscale defined by the dynamical heterogeneity is not static (one-time spatial correlation) but dynamic (two-time spatial correlation). 

Modern spin-glass-type theories of the structural glass transition~\cite{lubchenko2007, Biroli2008, Parisi2010} introduce a possible static length called `point-to-set' length that makes neither assumption nor description of the `amorphous order' underlying the glass transition. It was demonstrated theoretically~\cite{Montanari2006} for lattice models that the growth of a dynamical length scale imposes a lower bound on the point-to-set length. However this constraint is expected to become relevant only in a low temperature regime inaccessible to present-day simulations~\cite{Kob2011}. Indeed static lengths have been measured  uncorrelated to the dynamic one above this regime~\cite{Charbonneau}. The relation between the dynamical and the static length scales is a corner stone of the debate between geometrical frustration-based descriptions, facilitation-based descriptions, and the random first order transition (RFOT) theory.

In order to unveil static quantities, it is tempting to link dynamical heterogeneity to structural heterogeneity. Stable structures should move less than unstable ones. \citet{Widmer-Cooper2005} demonstrated that the initial configuration of a simulation, independently of the initial dynamics, has a role in the heterogeneity of the dynamics. \citet{Berthier2007} pointed out later that this influence does not exist at the particle level, but on larger scales: medium-range particle configuration can be statistically linked to medium-range dynamical heterogeneity. However these arguments are based on the iso-configurational ensemble (starting several simulations with different dynamics but with the same initial configuration), a method that has no experimental equivalent.

Given that a glass shows no long-range periodic order, structures different from a crystal, icosahedral~\cite{steinhardt1983boo,sadoc1999geometrical, tarjus2005fba} or amorphous order~\cite{lubchenko2007}, are most often pointed as responsible for the dynamical arrest. Since the pioneering work of Frank \cite{Frank1952}, icosahedral order is the archetypal model of amorphous structures~\citep{Spaepen2000}. Locally, an icosahedron maximises the density of packing, but its five-fold symmetry is incompatible with long-range periodicity. Geometrical frustration associated with icosahedral ordering is proposed to be crucial for vitrification ~\cite{steinhardt1983boo,sadoc1999geometrical,tarjus2005fba}. It was also proposed that the percolation of icosahedral order is responsible for glass transition~\cite{Tomida1995}.  Evidence of icosahedral ordering has indeed been found in very dense hard sphere packings~\citep{Bernal1960,Clarke1993, Anikeenko2007,Malshe2011,Charbonneau}, in various simulation models~\citep{steinhardt1983boo,Tomida1995,Doye2003,Pedersen2010,Coslovich2011} and more recently in experimental dense metallic liquids~\citep{Reichert2000,Celino2007} and glasses~\citep{Luo2004,Wang2011}.

Geometrical frustration in icosahedral order (or more generally locally favoured structures of the liquid) plays a central role in original versions of the spin-glass-type theory~\cite{steinhardt1983boo,sadoc1999geometrical} and the frustration-limited domain theory of the glass transition (see~\citep{tarjus2005fba} for a review). The former theory considers the icosahedral ordering under frustration. On the other hand, the latter theory takes as a reference state an unfrustrated icosahedral order existing in a curved space, even if (and because) this reference state is impossible for the real system to reach. If the space were curved, the locally preferred order of the liquid (icosahedral) would form a continuous ordered phase upon cooling. In the  euclidean space, however, this transition is avoided, yet the avoided transition temperature still acts as a critical point giving rise to diverging lengthscales (icosahedral domain size and defect size) that would explain the dynamical arrest. This theory suggests that the frustration-limited domains are the origin of the dynamical heterogeneity. Both theories are still very popular and icosahedral order is often cited as the reference amorphous structure particularly in metallic glasses~\citep{Reichert2000, Celino2007, Luo2004, Wang2011}. Yet, to our knowledge, numerical systems where a link between icosahedral order and dynamical heterogeneity has been observed~\citep{Doye2003, Pedersen2010, Coslovich2011}, always have a (quasi-)crystalline ground state including icosahedra. Thus, icosahedra could be regarded as a structural motif linked to the crystalline ground state in these systems. From this point of view, these results support neither the spin-glass nor frustration-limited domain approaches based on frustration in icosahedral order, but rather the crystal-based approach (see below).

An alternative way of explaining the phenomenology of the glass transition is based on the structure of the crystal. Crystallisation, even if avoided, may have an influence on the supercooled fluid or the glass~\citep{TanakaGJPCM, Cavagna2003, VanMegen2009a} and is taken as a reference state. Simulations~\cite{tanaka2010critical, Pedersen2010, Coslovich2011} and experiments~\citep{tanaka2010critical} show that a local bond orientational order compatible with the ground state (crystal) symmetry does extend to medium range in several model supercooled systems at accessible temperatures. In this class of systems, these transient but relatively stable structures seem to be correlated with the slow regions because they are low free-energy configurations. Furthermore, they lack extended positional order and thus could not be detected by macroscopic diffraction experiments. We emphasize that high crystal-like bond orientation regions are not crystal nuclei with solid nature, but transient medium-range structural order formed in a liquid as thermal fluctuations. 

In the present work, we focus on the colloidal (hard spheres) supercooled liquid \cite{pusey1987ogt} for two reasons. First, hard sphere-like colloidal particles can be tracked by confocal microscopy, giving access to the positions and dynamics of individual particles \cite{kegel2000swe, weeks2000}, thus allowing microscopic analysis of the dynamical and structural heterogeneity. Second, hard spheres have a well defined crystalline ground state of face-centred-cubic (\textsc{fcc}) or hexagonal-close-packed (\textsc{hcp}) symmetry and locally favoured structures of icosahedral symmetry. This enables us to distinguish the different scenarios of glass transition, in particular, the frustration-limited domain scenario, the amorphous order scenario, and the medium-range crystalline order scenario. We will show that both kinds of order are statistically associated with slower than average particles. However icosahedral order remains local, whereas the crystal-like order correlation length grows when approaching the glass transition. Moreover, the medium-range crystalline order is associated with the slow regions of the dynamic heterogeneities. We confirm that both structural and dynamical length scale grow in the same (critical-like) way. Our result suggest that, when not compatible with an avoided crystallisation, local ordering seems to play a minor role in the dynamical aspects of the glass transition.

\section*{Results}

\subsection*{Dynamics}

\begin{figure}
\begin{center}
\includegraphics{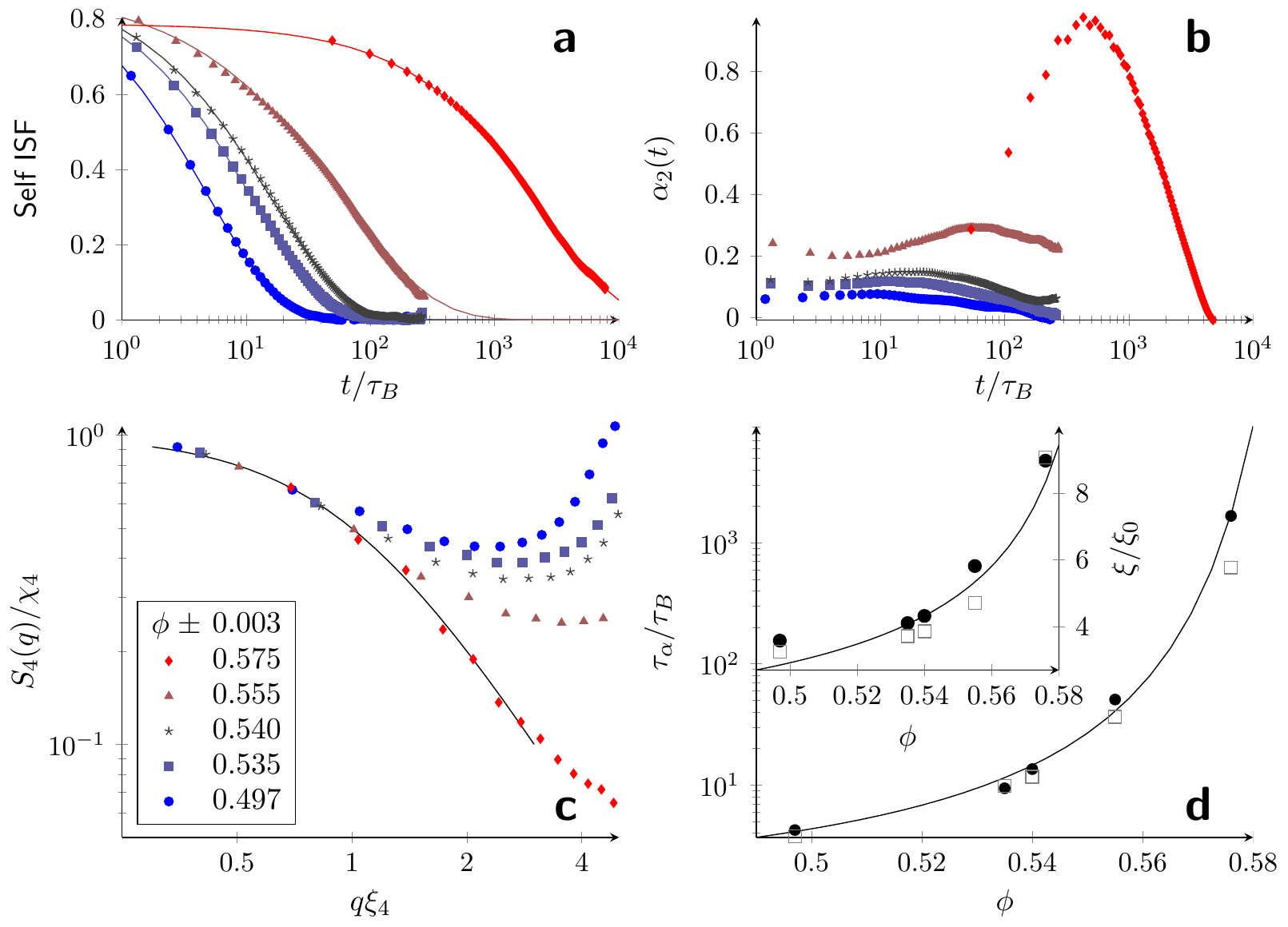}
\end{center}
\caption{\textbf{Dynamics of the system.} 
	\textbf{a,} Decay of the self-intermediate scattering (ISF) function computed at wave vector $q=2\pi/\left\langle \sigma\right\rangle$. Lines are stretched exponential fits from which we extract the structural relaxation time $\tau_\alpha$. 
	\textbf{b,} Non-Gaussian parameter, whose maximum defines the characteristic time of the dynamic heterogeneity $t^\text{dh}$.
	\textbf{c,} Collapse of the small-$q$ regimes of the four-point structure factor on the Orstein-Zernike function (solid curve).
	\textbf{d,} $\phi$-dependence of $\tau_\alpha$ (circles) and of $t^\text{dh}$ (squares). The solid curve is the VFT fit for $\tau_\alpha$ ($\phi_0=0.600$, $D=0.328$). 
	\textbf{Inset:} Dynamical ($\xi_4$, squares) and structural ($\xi_6$, circles) correlation lengths  scaled by their respective bare correlation length, respectively $\xi_{4,0}=0.206\sigma$ and $\xi_{6,0}=0.129\sigma$. The curve is the power-law given in the text.
	All times are scaled by the Brownian time $\tau_B$.}
	\label{fig:vft}
\end{figure}

We were able to follow the volume fraction dependence of the dynamics over nearly three orders of magnitude (see Fig.~\ref{fig:vft}), confirming the basic phenomenology observed in experiments~\citep{pusey1987ogt, kegel2000swe, weeks2000, BerthierR} as well as simulations~\citep{tanaka2010critical} of similar systems. In Fig.~\ref{fig:vft}a we fit the self-intermediate scattering function by a stretched exponential $\propto e^{-(t/\tau_\alpha)^\beta}$ that defines the structural relaxation time $\tau_\alpha$. 
Here $\beta$ is the stretching exponent. 

The non-Gaussianity of the dynamics is monitored by the kurtosis of displacements distribution (Fig.~\ref{fig:vft}b)
\begin{equation}
	\alpha_2(t) = \frac{3 \left\langle {(\Delta r(t))^4} \right\rangle}{5 {\left\langle {(\Delta r(t))^2} \right\rangle}^2}-1, 
	\label{eq:ng}
\end{equation}
where $\Delta r(t)= |\mbox{\boldmath$r$}(t_o+t)- \mbox{\boldmath$r$}(t_0)|$ is the displacement of a particle after a time interval $t$. At a given volume fraction, $\alpha_2(t)$ peaks at the characteristic time scale of the dynamic heterogeneity $t^\text{dh}(\phi)$.

The volume fraction dependence of $\tau_\alpha$ is super-Arrhenius and we fit it in Fig.~\ref{fig:vft}d by the Vogel-Fulcher-Tammann (VFT) law $\tau_\alpha=\tau_\alpha^0 \exp(D\phi/(\phi_0-\phi))$, where $D$ is the fragility index, $\phi_0$ is the hypothetical ideal glass transition volume fraction, and $\tau_\alpha^0$ is the microscopic timescale.  We notice that $t^\text{dh}$ corresponds to the end of the plateau, always slightly shorter than $\tau_\alpha$.

The spatial heterogeneity of the dynamics is characterised by a four-point susceptibility $\chi_4(t)$ and a four-point structure factor $S_4(q,t)$~\cite{Flenner2011}. To the reader unfamiliar with these terms (defined in Methods), we note that $S_4(q,t)\times N/N_s$ is the structure factor of a system containing only the slow particles.

As shown in Fig.~\ref{fig:vft}c, the small wave-vector behaviour of $S_4(q,t)$ is well described by the asymptotic Ornstein-Zernike function in Fourier space:
\begin{equation}
	S_4(q\rightarrow 0,t) \approx \frac{\chi_4(t)}{1+\xi_4(t)^2 q^2},
	\label{eq:OZ_Fourier}
\end{equation}
where $\xi_4(t)$ is the four-point correlation length (see Methods for the details of the fit). We found that at each volume fraction the maximum of $\chi_4(t)$ is reached at times longer than the relaxation time $\tau_\alpha$. In summary, we have
$t^\text{dh} \leq \tau_\alpha \leq t_{\chi_4}$.
For simplicity, we omit time dependence in $S_4$, $\xi_4$ and $\chi_4$ when $t=t_{\chi_4}$. The volume fraction dependence of the resulting correlation length is plotted in the inset of Fig.~\ref{fig:vft}. In the volume fraction range that we were able to study, it follows a power law
\begin{equation}
\xi_{4} =\xi_{4,0} ((\phi_0-\phi)/\phi_0)^{-\nu},
\label{eq:xi4ofphi}
\end{equation}
where $\nu=2/3$ and $\phi_0$ is independently determined by the above VFT fit; thus the bare correlation length $\xi_{4,0}$ is the only fitting parameter~\citep{tanaka2010critical}. Note that the lengths are in units of the mean diameter $\sigma$.

\subsection*{Bond orientational ordering}

\begin{figure}
\begin{center}
\includegraphics{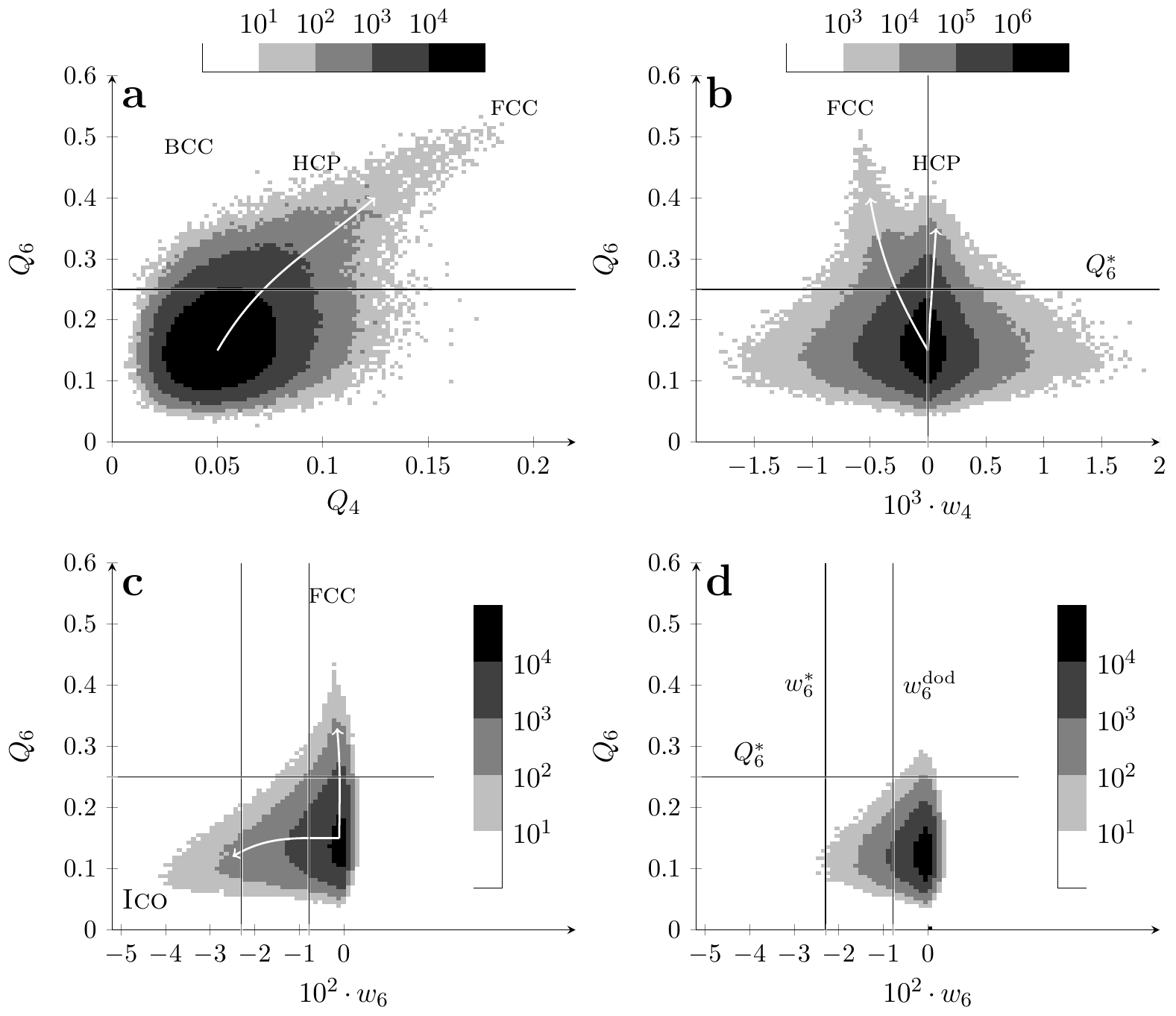}
\end{center}
\caption{\textbf{Population of local particle configurations specified by the }\textsc{boo}\textbf{ parameters}. The value of \textsc{boo} parameter for a perfect structure is  indicated by its name's position on each map. \textbf{a} in the $(Q_4,Q_6)$-plane, \textbf{b} $(w_4,Q_6)$-plane and \textbf{c-d} $(w_6,Q_6)$-plane. \textbf{a-c} are the populations at deep supercooling ($\phi=0.575\pm 0.03$); \textbf{d} is the same as \textbf{c} but for a liquid near the freezing point ($\phi = 0.497 \pm 0.03$). The colour at a given couple of BOO parameters indicate how often (log scale) this couple is detected (per units of area in this plane). The arrows stress the ordering tendencies: the tendency towards \textsc{fcc} is always visible, a weak tendency towards \textsc{hcp} can also be distinguished in \textbf{b}, and the tendency towards icosahedral order (\textsc{ico}) is visible in \textbf{c}. 
Conflicting tendencies towards \textsc{fcc} and \textsc{ico} in \textbf{c} is a manifestation of competing orderings, or frustration, between them. Straight lines are drawn at important thresholds: $Q_6^*$, $w_6^*$, $w_6^\text{dod}$ and $w_4=0$ that differentiates between \textsc{fcc} and \textsc{hcp}.}
	\label{fig:maps}
\end{figure}

The local structures of our suspensions are identified by a detailed analysis of Steinhardt's bond orientational order (\textsc{boo})~\citep{steinhardt1983boo}, including original improvements (see Methods). Figure~\ref{fig:maps} summarizes our structure identification results. We find the distinctive signature of medium-ranged crystalline order (\textsc{mrco}) of \textsc{fcc} type, without excluding some \textsc{hcp}. We find no trace of \textsc{bcc}-like structure. As for the aperiodic structures, icosahedral order stands out without any other obvious rivals (dodecahedra are present but can be considered as twisted icosahedra).

We establish that in our system the tendency towards icosahedral order is best monitored by the order parameter called $w_6$, whereas $Q_6$ is the natural crystal-like bond 
orientational order parameter (see Methods for the definitions of $w_6$ and $Q_6$). The population of local structures in the $(w_6,Q_6)$-plane is shown in Fig.~\ref{fig:maps}c-d, which confirms that the two tendencies are present in the supercooled liquid of hard spheres. In particular, icosahedra are present even at a relatively low volume fraction (Fig.~\ref{fig:maps}d). Both tendencies become more pronounced for deeper supercooling (Fig.~\ref{fig:maps}c). Furthermore, Fig.~\ref{fig:maps}c shows clearly that icosahedral order and crystalline order are incompatible and frustrate one another.

\subsection*{Bond order mobility}
In the introduction we mentioned the dynamic propensity invented by \citet{Widmer-Cooper2005} as a trick only accessible via simulations. In our system, since we know the relevant structural order parameters ($w_6$ and $Q_6$), we can compute the square displacement of all the particles with the same initial local structure (iso-bond order ensemble) and call it bond order mobility. We define the $w_6$-mobility as
\begin{equation}
	\Delta r^2(w_6, t) \equiv \left\langle \frac{
		\sum\limits_i{
			\left\|\mathbf{r}_i(t)-\mathbf{r}_i(0)\right\|^2 \delta(w_6^i-w_6)
			}
	}{
		\sum\limits_i{\delta(w_6^i-w_6)}
	}\right\rangle, 
	\label{eq:bo_propensity}
\end{equation}
where the brackets denote ensemble averaging. $\Delta r^2(Q_6, t)$ and $\Delta r^2(w_6, Q_6, t)$ can be defined in the same way. 
The iso-configurational propensity is very powerful since it does not require any {\it a priori} information on structures and thus can probe any static order if it exists. This feature is absent in the bond order mobility, which instead has the merit of accessibility even in a system that cannot be restarted ab libitum from exactly the same overall configuration. In general the two quantities are not equivalent.

Figure~\ref{fig:msd_Q6_w6} shows the bond order mobility (calculated at the characteristic time of the dynamical heterogeneity, $t^\text{dh}$). Both structural order parameters have a clear correlation with the dynamics: the more structured is the environment, the more likely the particle will be slow. For all volume fractions, the mobility displays the same linear dependence upon $w_6$ (Fig.~\ref{fig:msd_Q6_w6}b). Once scaled by the mean square displacement all points collapse on the same line. This is not the case for the $Q_6$ dependence (Fig.~\ref{fig:msd_Q6_w6}c) which is more and more pronounced with increasing the degree of supercooling. In our most deeply supercooled sample, the crystal-like environments can be in average ten times slower than the bulk, whereas even the almost perfect icosahedra are in average only $40\%$ slower than the bulk. A complete screening of all the possible \textsc{boo} mobilities excludes a link of slow dynamics to any other local symmetries (including exotic amorphous order), at least in our system.

To explain our mobility data, we can refer to the cage picture, where a particle is considered as trapped in the cage formed by its neighbours. Once the particle has escaped from its cage, it is free to diffuse. In general, this picture is not correct due to dynamical heterogeneity, or nonlocal cooperativity of motion. In the case of the icosahedral environment, however, it seems to hold: the normalisation by the bulk \textsc{msd} takes into account the diffusion once out of the cage, and the remaining universal dependence upon $w_6$ holds the information about the ``quality'' of the cage. Efficient packing makes the 13 particles icosahedra more stable than a disordered structure. Thus the central particle has a low probability to escape from its cage and start diffusing. After cage escaping, however, it diffuses in average like any other particle. This suggests that the influence of icosahedral order on the dynamics is only local because of its isolated character. By contrast the non-trivial $\phi$ dependence of the $Q_6$-mobility calls for nonlocal explanations.

\subsection*{Real space patterns and length scale}
\begin{figure}
\begin{center}
\includegraphics{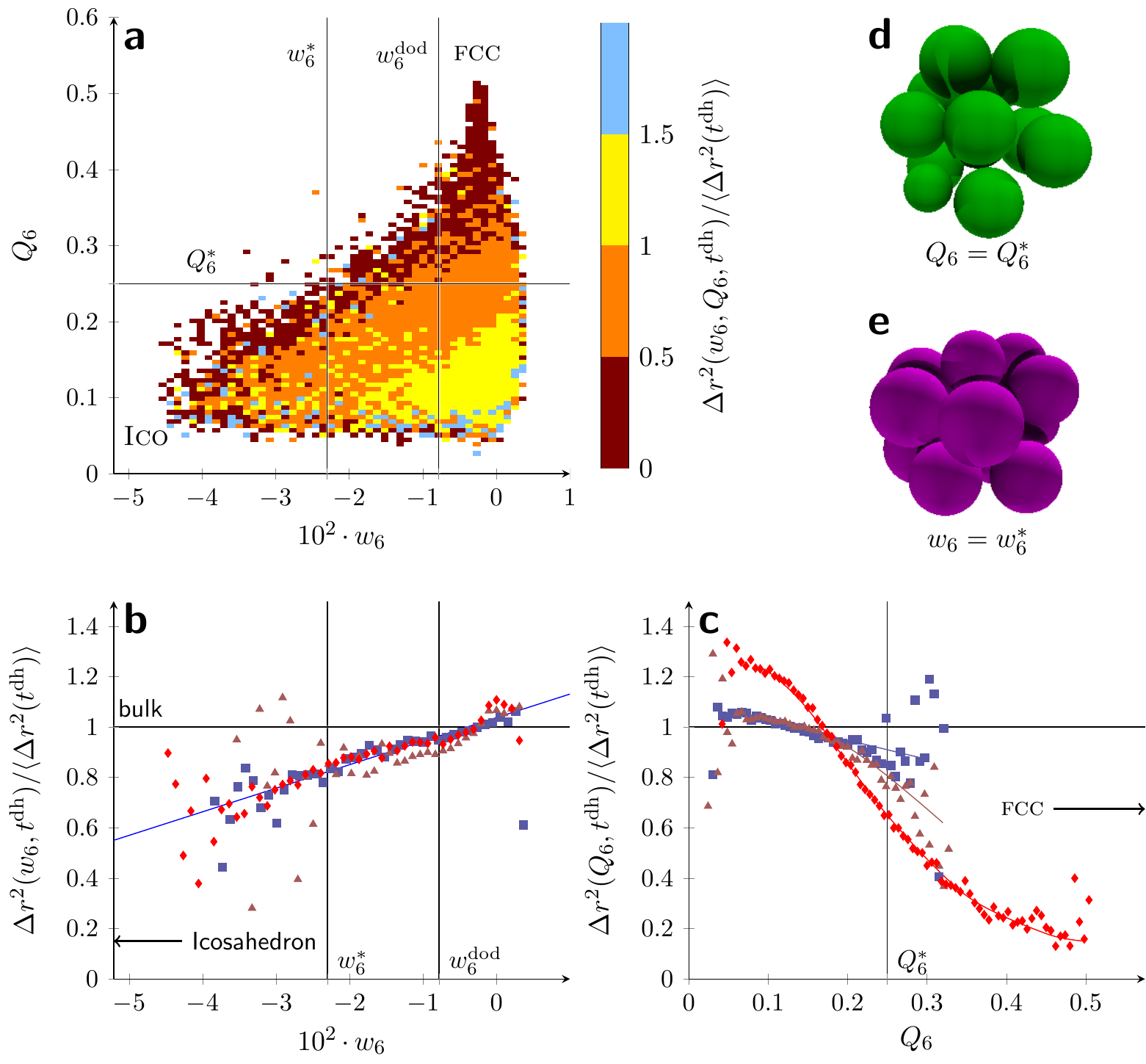}
\end{center}
\caption{\textbf{Bond order mobility.} {\bf a,} Normalised mobility in the $(w_6, Q_6)$-plane for our most deeply supercooled sample ($\phi = 0.575 \pm 0.03$). The colour scale is saturated at $1.5$ times the bulk mean square displacement. {\bf b-c,} Normalised mobility for icosahedral and crystalline order parameters respectively at volume fraction $0.535$ (squares), $0.555$ (triangles) and $0.575$ (diamonds), all $\pm 0.03$. Bulk mean square displacement is scaled to be at 1 (horizontal line). Perfect structures are on the edge of each plot. The lines are a guide for the eye, stressing the collapse of the $w_6$-mobility at all volume fractions in {\bf b} and the absence of such collapse in {\bf c}. The scattering at low volume fractions is due to poor averaging of rare structures. Straight lines in \textbf{a-c} corresponds to the important thresholds: $Q_6^*$, $w_6^*$ and $w_6^\text{dod}$. Examples of crystal-like cluster and distorted icosahedron at the respective threshold values are shown in \textbf{d} and \textbf{e}, respectively.}
	\label{fig:msd_Q6_w6}
\end{figure}

\begin{figure}
\includegraphics{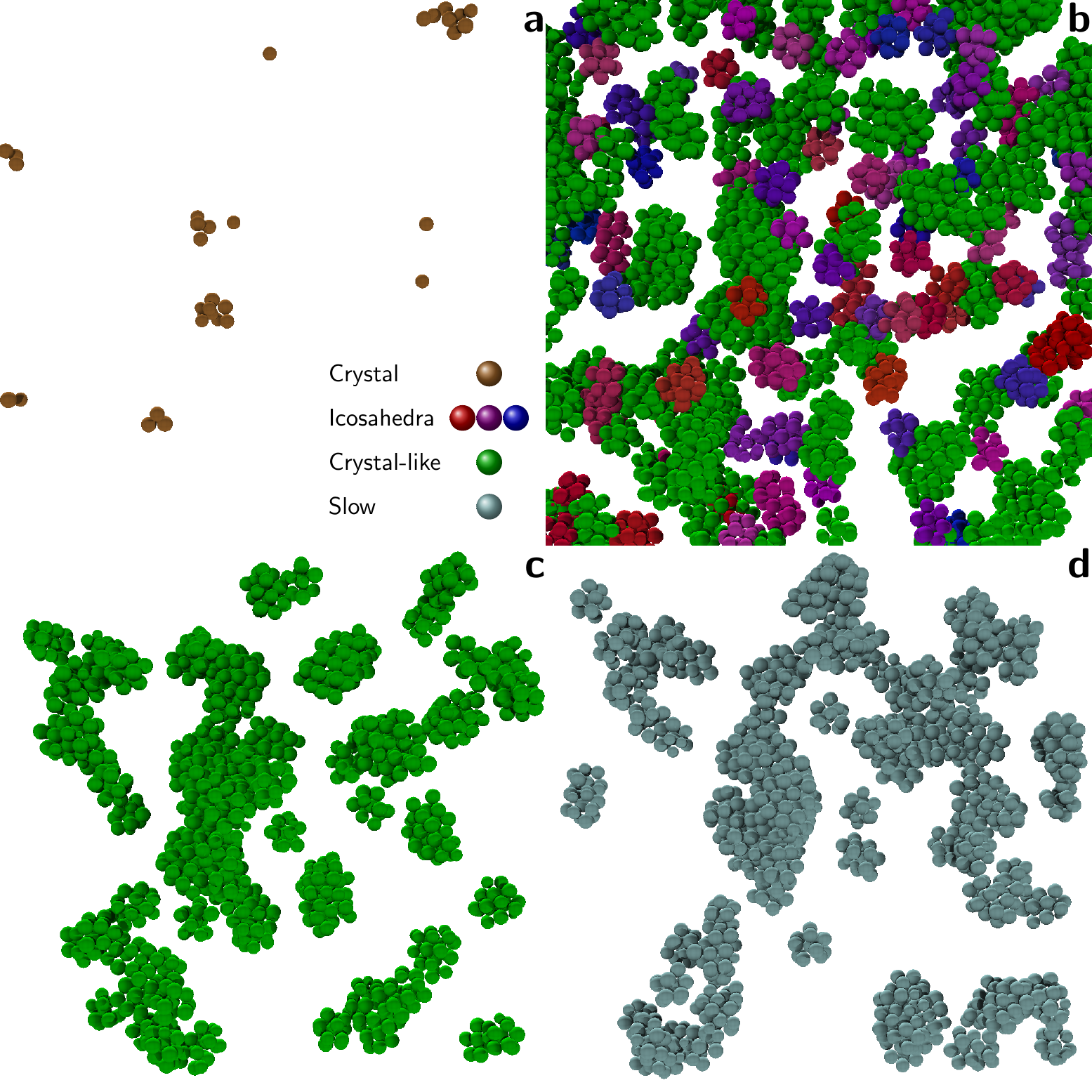}
\caption{\textbf{Computer reconstruction from confocal microscopy coordinates in our most deeply supercooled sample.} Depth is $\sim 12\sigma$. Only particles of interest and their neighbours are displayed. Each particle is plotted with its real radius. \textbf{a,} Particles with more than $7$ crystalline bonds. \textbf{b,} A typical configuration of bond ordered particles. Icosahedral particles are shown in the same colour if they belong to the same cluster. If a particle is neighbouring both crystal-like and icosahedral structures, it is displayed as icosahedral. \textbf{c,} Crystal-like particles alone (the order parameter was averaged over $t^\text{dh}/2$). \textbf{d,} Slow particles with respect to the coarse-grained displacement. Due to particles going in and out of the field of view, assignment of particles located very near the edges of \textbf{c} and \textbf{d} were not accurate and have been removed.}
	\label{fig:3D}
\end{figure}

To observe the ordered regions in real space, we define the thresholds $w_6^*$ and $Q_6^*$ so that the bond order mobility at the threshold is half between the bulk and the (extrapolated) perfect structure. In our most deeply supercooled sample, this criterion yields $Q_6^* = \Qstar$ for crystal-like structures (see Fig. ~\ref{fig:msd_Q6_w6}d) and $w_6^* = \wstar$ for icosahedral structures (see Fig.~\ref{fig:msd_Q6_w6}e). These thresholds are arbitrary but coherent with each other. We display in Fig.~\ref{fig:3D}b typical configurations of the ordered neighbourhoods in our deepest supercooled sample. With our thresholds, we see only small patches of icosahedral order that are not reaching medium range, whereas clusters of crystal-like order are much larger than icosahedral clusters and their sizes reach medium range. 

Here it may be worth stressing that \textsc{mrco} does not correspond to crystal nucleus (compare Fig.~\ref{fig:3D}a and b). However sub-critical nuclei are not unrelated to crystal-like \textsc{boo}. Nuclei are systematically embedded into much larger \textsc{mrco}. One can explain this phenomenon as perfect wetting of the nuclei by coherently bond-ordered regions \cite{Kawasaki2010c}. 
We note that the size of the crystal nuclei is smaller than the critical nucleus size and thus they appear only transiently. 
Furthermore, examples of crystal-like structures without embedded crystal nucleus, in addition to the continuity of $Q_6$ values (Fig.~\ref{fig:maps}), suggest that the crystal nuclei are the extreme part of the bond order fluctuations. 
Indeed it has been shown in hard spheres~\cite{OMalley2005, Schilling2010, Kawasaki2010c}, Gaussian-core model~\cite{Russo2012} and probably Wahnstr\"om binary Lennard-Jones~\cite{Pedersen2010} that parts of \textsc{mrco}, fluid regions with a structure reminiscent of the crystal, act as precursors to crystallisation if the local density there is high enough. However in these systems the presence of precursors to crystallisation does not imply the formation of a critical nucleus and growth. Furthermore, \textsc{mrco} is always present (transiently) in the supercooled liquid: \textsc{mrco} are usual fluctuations of the system, as common as the fast and slow regions of the dynamic heterogeneity~\cite{tanaka2010critical}. We also note that the average density within \textsc{mrco} is the same as within disordered liquid regions and thus different from within crystal nuclei~\cite{Kawasaki2010c}.

Now we consider a link between static spatial structures and dynamics. 
Reflecting the short-range nature of icosahedral order, we were not able to extract the associated lengthscale that would grow when approaching the glass transition. By contrast, the spatial extent of crystal-like \textsc{boo} is well described by $G_6(r)$, the spatial correlation function of $Q_{6 m}$ (see, e.g., \cite{tanaka2010critical}). Both dynamical and structural lengthscales (defined in Methods) are increasing while approaching the glass transition in a coherent manner (the inset of Fig.~\ref{fig:vft}). We can see that near $\phi_0$ the growth of both dynamic and static correlation length can be well expressed by the same power law (Eq.~(\ref{eq:xi4ofphi})), although the range is rather limited. In agreement with~\citep{tanaka2010critical} it suggests that the dynamical heterogeneity in hard spheres is a manifestation of critical-like fluctuations of the crystal-like bond orientational order parameter, and neither those of icosahedral nor amorphous order. In relation to this, we stress that $G_6(r)$ is not sensitive to aperiodic structures such as icosahedral and amorphous order (see Methods). 

Of course, one can set a more permissive threshold on $w_6$ and observe more icosahedra. A percolating network of icosahedral neighbourhood can even be found at a high volume fraction if the threshold encompasses enough particles (see Fig.~\ref{fig:percolation}c). However, the imperfect icosahedra particles that must be included to form this network have a mobility comparable to the bulk (Fig.~\ref{fig:msd_Q6_w6}b). Furthermore, we checked that this percolation is of the 3D random percolation class (cluster size distribution follows a power-law of exponent $\approx 2.1$, see Fig.~\ref{fig:msd_Q6_w6}a), meaning that a comparable network could be obtained by picking randomly the same number of particles (see Fig.~\ref{fig:percolation}d). We conclude that the concept of icosahedral network is not physically meaningful at least in our system.

As a final but an intuitive check, we compare the spatial repartition of stable crystal-like order (Fig.~\ref{fig:3D}c) and slow particles (Fig.~\ref{fig:3D}d). As shown in~\citep{Berthier2007}, the correlation between structure and dynamics is not a one-to-one correspondence at the particle level but a statistical relationship on larger length scales. We were able to show visually this correlation by averaging out short time ($t^\text{dh}/2$) fluctuations of $Q_6$; by calculating the square displacement of each particle over $2\tau_\alpha\approx 4.5t^\text{dh}$ which allow to accumulate many rearrangement events; and by coarse-graining this value over the particle's neighbours~\citep{Berthier2007} (see Eq.~(\ref{eq:Mu}) in Methods). Figure~\ref{fig:3D}d displays the $10\%$ slowest particles according to this criterion and their neighbours. The correlation of shape and size is clearly seen between Fig.~\ref{fig:3D}c and Fig.~\ref{fig:3D}d, although it is not perfect. The slight disagreement comes from the following reasons: some clusters of \textsc{fcc}-like order disappear soon after the initiation of the displacement measurement and some particle groups, which come back to their original positions after the period of $2\tau_\alpha$, can be accounted as slow particles. We could find no such correlation between icosahedral particles (Fig.~\ref{fig:3D}b) and slow regions.

Note that we recover the correlation between crystal-like order and slowness by two independent methods: directly in instantaneous \textsc{boo} space via mobility on moderate times; and indirectly via positions and shapes of (short time averaged) ordered regions coherent with those of slowest regions (on long times). In the former method, quality of statistics comes from ensemble averaging over many realisations of each structure. In the latter method, it comes from both coarse graining and the length of trajectories that allow to accumulate many rearrangement events.

\section*{Discussion}
We have demonstrated that supercooled hard-sphere liquids display local icosahedral ordering as well as structural ordering linked to the avoided crystalline (\textsc{fcc}) ground state. Both types of packing have low local free energy and thus are locally favoured. Yet the icosahedral order remains local even at deep supercooling and (consistently with the arguments of~\citep{Berthier2007}) is not able to influence significantly the global dynamics. By contrast, the slowing down associated with the crystal-like \textsc{boo} is nonlocal and long-lived. The characteristic length scale of the crystal-like order grows when approaching the glass transition in the same way as that of the dynamical heterogeneity. This suggests that the structural origin of the dynamical arrest is linked to the avoided crystallisation and neither to the condensation of the local (icosahedral) order of the liquid nor to the exotic amorphous order at least in our system. We note that the development of structural order and its link to dynamic heterogeneity are at odds with purely kinetic scenarios of glass transition.

How general is our crystal-like order scenario? Not all glass formers crystallise, indeed the crystal structure of many glass formers is unknown. However in a system where the crystallisation is possible but unlikely (a definition that include many practical glass formers), we expect structures locally reminiscent of the symmetry of the crystal to be present as common fluctuations of the supercooled liquid. Their relative stability and ability to reach medium-range sizes makes them a good candidate to explain the slow regions of the dynamic heterogeneity. In that class of systems static and dynamical lengths would grow together. Yet the generality and limitation of this scenario need to be checked carefully in the future.

We note that \citet{Charbonneau} are not finding growing 6-fold bond order length scale in a binary hard sphere mixture. \citet{Hopkins2011b,Hopkins2012} showed that for this size ratio if the two components de-mix they crystallise independently in \textsc{fcc} or \textsc{hcp}; however the symmetry-breaking phase of the mixed system is not known and could possibly be not 6-fold symmetric. This de-mixing required for crystallization may destroy a link between configurations of low local free energy in the liquid state and the crystal structure. Indeed, the study of \textsc{boo} in binary systems with the concept of bond orientational order is not promising, as reported in~\citep{tanaka2010critical, KawasakiJPCM} for 2D binary soft disk simulations.

For some systems, the first order transition avoided upon supercooling leads to a symmetry compatible with icosahedral order. The symmetry-breaking phase can be a quasi-crystal~\citep{Doye2003} or a Frank-Kasper crystal, as recently demonstrated for the Wahnstr\"om mixture~\citep{Pedersen2010}. In that case, the icosahedral order can be regarded as part of the defective crystalline bond orientational order that slows down the system, so the icosahedral order can play a role in the slowing down. However, we showed here that when the icosahedral order is not compatible with the crystal symmetry, its role in the slowing down is actually minor.

We do not say that the local (icosahedral) order of the liquid is unimportant in the glass transition. Actually it plays a role in preventing the first order transition (crystallisation) to happen, thus allowing supercooling and the glass transition~\cite{TanakaMJPCM}. Furthermore, this frustration against crystallisation may be linked to the fragility of the glass former~\citep{TanakaGJPCM,tanaka2010critical}. Competing crystal-like and non-crystal-like orderings may be important in many glass-forming systems covering colloidal, molecular, polymeric, oxide, chalcogenide, and metallic glasses. Indeed, our competing ordering scenario seems very consistent with a recent finding in metallic liquids~\cite{liu2010metallic} and may explain the high glass-forming ability of bulk metallic glassformers~\cite{Wang2004}. An intimate link between the liquid and crystal structure may also provide a mechanism of fast phase change materials \cite{wuttig2007phase}.

\section*{Methods}

\subsection*{Experimental}

We used \textsc{pmma} (poly(methyl methacrylate)) colloids sterically stabilized with methacryloxypropyl terminated \textsc{pdms}(poly(dimethyl siloxane)) and fluorescently labelled with rhodamine isothiocyanate chemically bonded to the \textsc{pmma}. The size distribution, as characterised by scanning electron microscopy imaging (see Supplementary Figure~S1), showed a long tail towards small sizes, leading to the polydispersity larger than 6\%. This amount of polydispersity allows us to avoid or at least delay crystallisation~\cite{Zaccarelli2009} (see below the crystallization monitoring method) but is too low for fractionation to happen~\citep{Fasolo2003}. The colloids were suspended in a solvent mixture of cis-decalin and cyclohexyl-bromide for both optical index and density matching. To screen any (weak) electrostatic interactions, we dissolved tetrabutylammonium bromide salt, to a concentration of \unit{300}{\nano\mole\per\liter}~\citep{royall2005}. The estimated Debye screening length is \unit{13}{\nano\metre}, well below the length scale of the colloids. To avoid errors due to the swelling of the particles, their size was measured in the same solvent by the following technique: short-ranged depletion attraction was induced by adding non-adsorbing polystyrene to a dilute suspension to drive particles into contact; the position of the first peak of the $g(r)$ gave the diameter $\sigma = \unit{3.356\pm0.03}{\micro\meter}$.

After careful shear melting, the samples are filled into square $\unit{500}{\micro\metre}$ capillaries (Vitrocom). The most deeply supercooled sample showed a slight indication of sedimentation even after careful density matching. To avoid this effect we used a thinner $\unit{100}{\micro\metre}\times\unit{1}{\milli\metre}$ cell. The data were collected on a Leica SP5 confocal microscope, using \unit{532}{\nano\meter} laser excitation. The temperature was controlled on both stage and objective lens, allowing a more precise density matching. Our polydisperse colloids were detected using a novel multi-scale algorithm (see Supplementary Methods on the details), which allows us to retrieve both position and size of each particle with about $1\%$ precision, and this for any arbitrary size distribution. We stress that standard particle tracking methods~\cite{kegel2000swe, weeks2000} are not able to track all the particles in our polydisperse samples (see Supplementary Figure~S2). The large size of our colloids allows a comparable precision in position for both in-plane and out-of-plane coordinates (see Supplementary Figure~S1). The particles are tracked in time to extract dynamical informations. Typically $\approx 5000$ trajectories can be followed over a few structural relaxation times.

\subsection*{Dynamics}

To characterise the dynamic heterogeneity, we define a microscopic overlap function $w_i(t) = \Theta [b - |\mathbf{r}_i(t) - \mathbf{r}_i(0)|]$, where $\Theta(x)$ is Heaviside’s step function, $\mathbf{r}_i(t)$ is the position
of particle $i$ at a time $t$ and $b=0.3\langle\sigma\rangle$.
The number of overlapping particles is $N_s(t)$ and in the thermodynamic limit its fluctuations define the four-point susceptibility:
\begin{equation}
\chi_4 = (\langle N_s^2\rangle - \langle N_s\rangle^2) /  \langle N\rangle. 
\label{eq:chi_4}
\end{equation}
In a simulation with a fixed number of particles $N$ and a limited size, some fluctuations are forbidden, thus Eq.~(\ref{eq:chi_4}) do not give the right value of $\chi_4$ (see \citep{Flenner2011}). However in our experiments we are observing a small portion of the whole colloidal suspension, thus we can use time average to sample all the fluctuations of this large system, and Eq.~(\ref{eq:chi_4}) becomes exact and ensemble-independent.

The four-point structure factor~\cite{Flenner2011} is
\begin{equation}
	S_4(q,t) = N^{-1}(\left\langle W(\mathbf{q},t) W(-\mathbf{q},t) \right\rangle - | \left\langle W(\mathbf{q},t) \right\rangle^2 |),   
	\label{eq:S4}
\end{equation}
where $W(\mathbf{q},t)$ is the Fourier transform of $w_i(t)$: 
\begin{equation}
	W(\mathbf{q},t) = \sum_i w_i(t)\exp(-\imath \mathbf{q}\cdot\mathbf{r}_i(0)).  
\end{equation}

Our experimental data do not have periodic boundary conditions, so we must use a window function to ensure the correct correlation, especially at small $q$. Here we use the Hamming window, but we checked that our results are not affected by other reasonable choices of the window function. The results are shown in Fig.~\ref{fig:vft}c.

As stressed in \citep{Flenner2011}, the value $S_4(q=0)$ is crucial for the quality of the fit in Eq.~(\ref{eq:OZ_Fourier}). Here we use the value of $\chi_4$ given by Eq.~(\ref{eq:chi_4}) and we fit $S_4(q)$ for $q<1.5/\xi_4$.

\subsection*{Structure identification}

\citet{steinhardt1983boo} defined the \textsc{boo} of the $\ell$-fold symmetry as a $2\ell+1$ vector:
\begin{equation}
	q_{\ell m}(i) = \frac{1}{N_i}\sum_{0}^{N_i} Y_{\ell m}(\theta(\mathbf{r}_{ij}),\phi(\mathbf{r}_{ij})),
	\label{eq:qlm}
\end{equation}
where the $Y_{\ell m}$ are spherical harmonics and $N_i$ is the number of bonds of particle $i$. In the analysis, one uses the rotational invariants defined as:
\begin{align}
	q_\ell =& \sqrt{\frac{4\pi}{2l+1} \sum_{m=-\ell}^{\ell} |q_{\ell m}|^2 }, \label{eq:ql}\\
	w_\ell =& \sum_{m_1+m_2+m_3=0} 
			\left( \begin{array}{ccc}
				\ell & \ell & \ell \\
				m_1 & m_2 & m_3 
			\end{array} \right)
			q_{\ell m_1} q_{\ell m_2} q_{\ell m_3}, \label{eq:wl}\\
	\hat{w}_\ell =& w_\ell{\left( \sum_{m=-\ell}^{\ell} |q_{\ell m}|^2 \right)}^{-\frac{3}{2}},
\end{align}
where the term in brackets in Eq.~(\ref{eq:wl}) is the Wigner 3-j symbol. Note that the popular~\citep{steinhardt1983boo,Lechner2008} $\hat{w}_\ell$ can show a large norm if its denominator (or $q_\ell$) has small values (particle with low $\ell$-fold symmetry). Therefore, we have used here $w_\ell$ that pulls clearly apart highly symmetric particles from the fluid distribution.

Following~\citet{Lechner2008}, we coarse-grain the tensorial \textsc{boo} over the neighbours:
\begin{equation}
	Q_{\ell m}(i) = \frac{1}{N_i+1}\left( q_{\ell m}(i) +  \sum_{j=0}^{N_i} q_{\ell m}(j)\right), 
	\label{eq:Qlm}
\end{equation}
and define coarse-grained invariants $Q_\ell$ and $W_\ell$ in the same way as Eqs.~(\ref{eq:ql}) and (\ref{eq:wl}). Structures with and without spatial extendability are then much easier to tell apart~\citep{Lechner2008}. For non-extendable local structures like icosahedra, their $Q_\ell$ and $W_\ell$ are buried into the liquid distribution. With this in mind, we looked for extendable structures in the $\lbrace Q_\ell, W_\ell\rbrace$ space, and for non-extendable (aperiodic) structures in the $\lbrace q_\ell, w_\ell\rbrace$ space, with $\ell=4,6,8,10$.

In both Eqs.~(\ref{eq:qlm}) and (\ref{eq:Qlm}), the sum runs over the ``neighbours'' (labelled with $j$) of particle $i$. The definition of the neighbours is crucial for the consistency of the analysis. The most popular criterion in literature is that particles closer than $R_\text{max}$ are neighbours, with $R_\text{max}$ the first minimum of $g(r)$. To account for the polydispersity, we used a threshold in the scaled distance $\hat{r}_{i j} = r_{i j} /(\sigma_i+\sigma_j) < \hat{R}_\text{max}$, with $\hat{R}_\text{max}$ corresponding to the first minimum of $g(\hat{r})$. We made a first round of analysis using this definition, constructing maps similar to Fig.~\ref{fig:maps} but noisier. These maps allowed us to discard the possibility of \textsc{bcc} (high $Q_6$, low $Q_4$, high $Q_{10}$) and generally any structure except \textsc{fcc} (high $Q_6$, high $Q_4$ and $w_4<0$), \textsc{hcp} (high $Q_6$ and $w_4>0$) or icosahedron (very negative $w_6$, high $q_6$ and $q_{10}$). Dodecahedron can be found for $w_6>w_6^\text{dod}\approx -0.00782$, i.e., between the liquid and the icosahedral order. We are thus left with structures where the particles naturally have 12 neighbours. 

We then took a new \textsc{boo} analysis, this time taking the first 12 closest particles (in $\hat{r}$) as neighbours. This criterion reduces the noise in the distribution of the invariants and makes detected structures more easily comparable with crystallographic archetypes. It also avoids artefacts near $R_\text{max}$ in the correlation functions and removes the negative correlation between the number of neighbours and the value of the $q_\ell$. The figures presented here were constructed from this second round of analysis.

\subsection*{Crystal nuclei}

To avoid any confusion between our imperfect crystal-like order and well-formed but small transient crystals, we detect crystal nuclei via the most standard method in the literature~\cite{ReintenWolde1996, Zaccarelli2009}. For each neighbouring particle, we compute the normalized scalar product of the (non coarse-grained) $q_{6 m}$:
\begin{equation}
	s(i,j) = \frac{
		\sum_{m=-6}^{6} q_{6 m}(i) q_{6 m}^{*}(j)
	}{
		\sqrt{\sum_{m=-6}^{6} |q_{6 m}(i)|^2} \sqrt{\sum_{m=-6}^{6} |q_{6 m}(j)|^2}
	}.
	\label{eq:boo_dot_product}
\end{equation}
The bond between particles $i$ and $j$ is considered crystalline if $s(i,j)>0.7$~\cite{Zaccarelli2009}. A particle is crystalline if more than half of its bonds ($\geq 7$) are crystalline.

Polydispersity is an important controlling factor which determines the ease of crystallization of colloidal suspensions (see, e.g., \cite{Zaccarelli2009}). 
Polydispersity of our sample was high enough to prevent crystallisation in bulk over our experimental time scale. 
However, heterogeneous nucleation from walls sometimes interfered our experiments, thus we had to discard such samples. 
We find no large crystal nucleus in the samples analysed in the text. Even at deep supercooling a crystal nucleus never gathers more than 8 particles. This is smaller than the critical nucleus size (2-3 particle diameters according to~\cite{Auer2001} and classical nucleation theory). In agreement, crystal-like solid regions appear only transiently without growth, i.e. we never observed crystal nucleation in bulk for our samples. Truly crystalline particles account for less than $1\%$ of the system. Figure~\ref{fig:3D} allows to compare the spatial extent of crystal nuclei with that of the crystal-like clusters or of the slow regions. It demonstrates that trivial crystallisation can be responsible for neither the spatial extent of the dynamical heterogeneities nor glassy slow dynamics.

\subsection*{Slow particles patterns}

The above definition of neighbours is used consistently in the coarse graining of square displacements:
\begin{equation}
	\bar{\Delta r^2}_i(t) = \frac{1}{N(i)+1}\left( {\Delta r^2}_i(t) + \sum_{j=0}^{N(i)} {\Delta r^2}_j(t) \right),
	\label{eq:Mu}
\end{equation}
where ${\Delta r^2}_i(t) = \left\|\mathbf{r}_i(t)-\mathbf{r}_i(0)\right\|^2$ and the neighbours are defined at initial time. 

\subsection*{Icosahedral percolation}

\begin{figure}
\begin{center}
\includegraphics{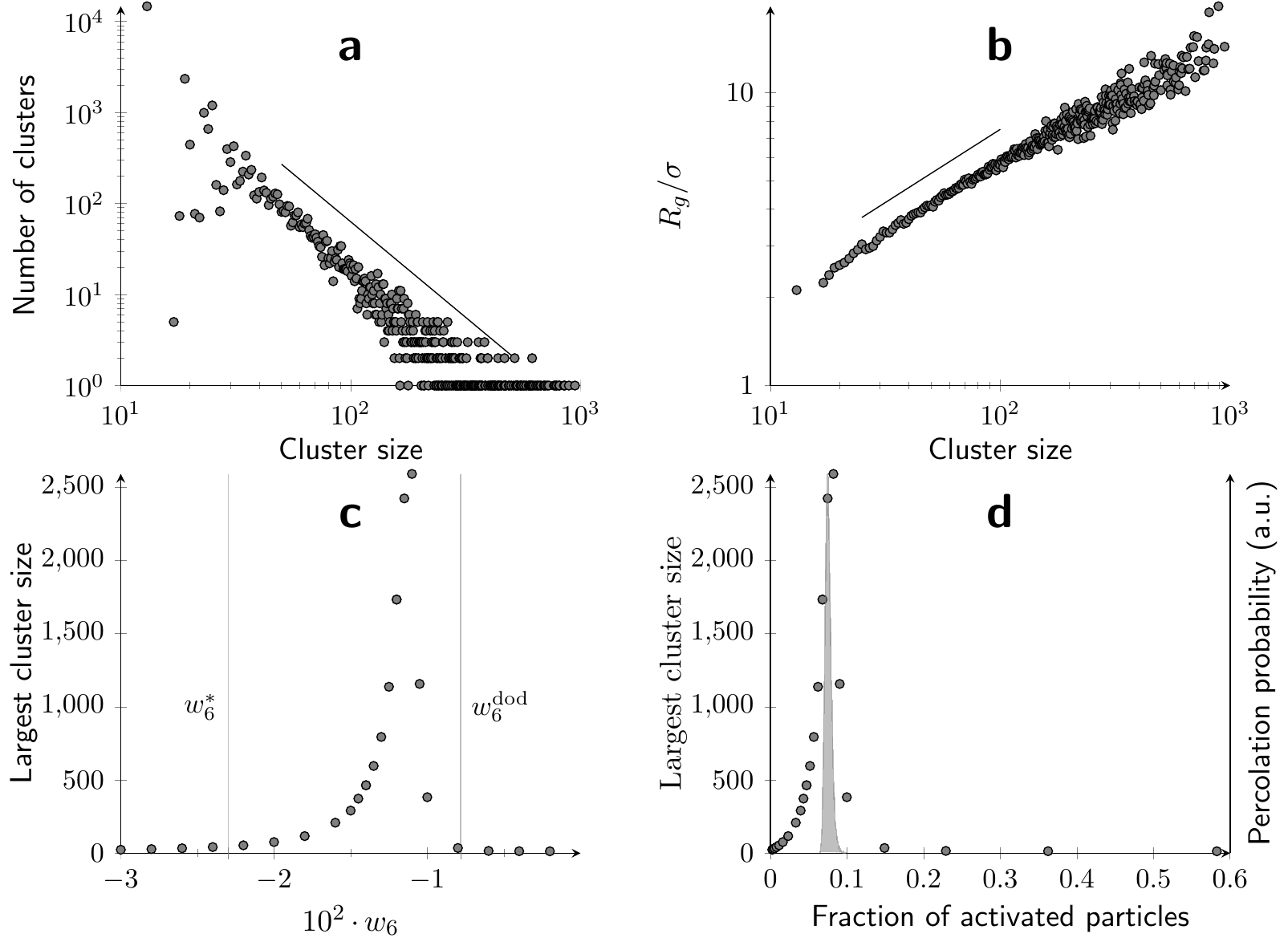}
\end{center}
	\caption{\textbf{Imperfect icosahedral clusters.} \textbf{a,} Cluster size distribution (compatible with random percolation) for nearly percolating threshold ($w_6<-0.012$). 
\textbf{b,} Radius of gyration (indicating a fractal dimension $d_f\approx 2$ given by the straight line) for the case of {\bf a}. \textbf{c,} Size of the largest non-percolating cluster as a function of the threshold on $w_6$. Percolation takes place near $w_6\approx -0.011$. Vertical lines indicate the position of $w_6^*$ and $w_6^\text{dod}$ respectively. \textbf{d,} Translation of \textbf{c} in terms of the fraction of activated particles (dots) together with the probability of the onset of percolation when particles are randomly activated (gray shading). We can see that activating particles as a function of icosahedral order and those activated randomly have almost the same percolation threshold ($\approx7.5\%$ of the particles activated).}
	\label{fig:percolation}
\end{figure}

We studied clusters formed by icosahedral particles as a bidirectional graph. The nodes of the graph are all the particles, and at the beginning we have no edge. When a particle is `activated' the bonds between this particle and its 12 neighbours are added to the graph. It is often meaningful to activate the particles as a function of an order parameter, for example sorted by increasing $w_6$; the most icosahedral particle connects first and we activate particles that are less and less icosahedral until percolation and beyond. For our most supercooled sample, the percolation threshold is identified near $w_6\approx -0.011$ (see Fig.~\ref{fig:percolation}c) which corresponds to $7.5\%$ of activated particles. We stress that this level of icosahedral order corresponds to an average mobility very close ($\gtrsim 90\%$) to the bulk (see Fig.~\ref{fig:msd_Q6_w6}b).

In this way, the cluster size distribution and fractal dimension can be studied systematically at each value of the threshold on the order parameter. In Fig.~\ref{fig:percolation}a-b we report characteristics of icosahedral clusters when the threshold is set just below percolation (all particles with $w_6<-0.012$ are activated). The cluster size distribution shows an exponent of $2.1$, characteristic of the random percolation universality class. The fractal dimension is close to $2$.

One can also activate particles in a random fashion and measure the fraction of activated particles needed for percolation to occurs. We performed 100 try of this Metropolis algorithm on 160 configurations, in order to estimate the percolation probability function of the fraction of activated particles. Results are reported in Fig.~\ref{fig:percolation}d and indicate that the peculiar case of icosahedral order clusters has nothing special.

\subsection*{Bond order spatial correlation}

We defined the coarse-grained bond order correlation function
\begin{equation}
	G_6(r) = \frac{4\pi}{13}\frac{\sum_{i,j} \sum_{m=-6}^{6} Q_{6 m}(i) Q_{6 m}^{*}(j) \delta(r_{ij}-r)}{\sum_{i,j} \delta(r_{ij}-r)},
	\label{eq:G_6}
\end{equation}
which is not sensible to aperiodic structures. After baseline subtraction, the envelope of $G_6(r)$ is fitted by the Ornstein-Zernike form to extract the crystal-like bond orientational order correlation length $\xi_6$. We checked that $g_6(r)$, defined from the non-coarse-grained bond order in the same way as Eq.~(\ref{eq:G_6}), yield similar results. Insensibility against aperiodic structures comes from the scalar product in Eq.~(\ref{eq:G_6})~\cite{Tomida1995}, not from coarse-graining.

\section*{}
\paragraph*{\bf Acknowledgments}
The authors thank Paddy Royall and John Russo for fruitful discussions. 
This work was partially supported by a grant-in-aid from the 
Ministry of Education, Culture, Sports, Science and Technology, Japan and 
Aihara Project, the FIRST program from JSPS, initiated by CSTP. 

\paragraph*{\bf Author Contributions}
H.T. conceived the research, M.L. performed experiments and analysis, M.L. and H.T. discussed and wrote the manuscript.

\paragraph*{\bf Author Information} 
The authors declare that they have no competing financial interests. 
Correspondence and requests for materials should be addressed to 
H.T. 

\clearpage

\clearpage

\clearpage

\clearpage

\clearpage

\end{document}